\documentclass[12pt]{iopart}
\usepackage{iopams}

\begin{document}

\title[Chaotic scattering with direct processes: A generalization of Poisson's kernel]{Chaotic scattering with direct processes: A generalization of Poisson's kernel for non-unitary scattering matrices}

\author{V. A. Gopar$^{1,2}$, M. Mart\'{\i}nez-Mares$^3$, and R. A. M\'endez-S\'anchez$^4$}
\address{$^1$ Departamento de F\'isica Te\'orica, Facultad de Ciencias, 
Universidad de Zaragoza, Pedro Cerbuna 12, 
50009 Zaragoza, Spain.}
\address{$^2$ Instituto de Biocomputaci\'on y 
F\'{\i}sica de los Sistemas Complejos, Universidad de Zaragoza,
Corona de Arag\'on, 42, 50009 Zaragoza, Spain.}

\address{$^3$ Departamento de F\'{\i}sica, Universidad Aut\'onoma
Metropolitana-Iztapalapa, A. P. 55-534, 09340 M\'exico D. F., Mexico.}
\address{$^4$ Instituto de Ciencias F\'{\i}sicas, Universidad Nacional Aut\'onoma de M\'exico, A.P. 48-3, 62210, Cuernavaca, Morelos, Mexico.}

\date{\today}

\begin{abstract}

The problem of chaotic scattering in the presence of direct processes or prompt responses is mapped via a transformation to the  case of scattering in the absence of such processes for {\it non-unitary} scattering matrices $\tilde S$. 
When prompt responses are absent, 
$\tilde S$ is uniformly distributed according to its invariant measure in the space of $\tilde S$ matrices with zero average $\langle \tilde S\rangle=0$. When direct processes occur, the distribution of $\tilde S$ is non-uniform and is characterised by an average $\langle \tilde S \rangle \neq 0$. In contrast to the case of unitary matrices $S$, where the invariant measures of $S$ for chaotic scattering with and without direct processes are related through the well-known Poisson kernel, we show that for non-unitary scattering matrices the invariant measures  are related by the Poisson kernel {\it squared}. 
Our results are relevant to situations where flux conservation is not satisfied, for  transport experiments in chaotic systems where gains or losses are present, for example in microwave chaotic cavities or graphs, and acoustic or elastic resonators.
\end{abstract}


\maketitle

\section{\label{sec:introduction}Introduction}

The statistical properties of ensembles of scattering matrices have been
studied extensively since their introduction in pioneering works in the field of
nuclear physics~\cite{porter,metha-book}. More recently, Random Matrix
Theory (RMT) techniques have been applied to study several statistical 
properties of electronic transport in mesoscopic systems~\cite{pier-book,beenakker-review,Alhassid}. 
Although the spectrum of applications of RMT is quite wide, ranging from atomic nuclei to microwave cavities, most of the investigations in RMT have considered systems in which flux is conserved. Therefore, the associated $n\times n$ scattering matrix $S$ is unitary, 
\begin{equation}
\label{eq:flux}
SS^{\dagger} = I_n, 
\end{equation}
where $I_n$ denotes the unit matrix of dimension $n$, and the dagger means Hermitian conjugation. Here $n$ denotes the number of scattering channels.

When unitarity is the only constraint, the unitary case denoted by $\beta=2$ 
in Dyson's scheme applies~\cite{Dyson}. If in addition time-reversal symmetry is imposed, then $S$ is also a symmetric matrix,
\begin{equation}
S = S^{T} ,
\end{equation}
a case corresponding to $\beta=1$. Here, the superscript $T$ denotes transposition. When time-reversal symmetry is present, but spin-rotation symmetry is broken, the $S$ matrix is unitary and self-dual, a case denoted by $\beta=4$.

Any scattering matrix $S$ can be decomposed as 
\begin{equation}
\label{eq:paraS}
S = U V, 
\end{equation}
where $U$ and $V$ are $n\times n$ unitary matrices for $\beta=2$, while $V=U^T$ for $\beta=1$. For $\beta=4$, $V=U^R$, where $U^R$ is the dual matrix of $U$. 

It has been shown~\cite{pier-book,beenakker-review} that chaotic scattering in the absence of direct processes is well described by uniformly distributed $S$-matrices. The uniform distribution is given by the invariant measure defined through the relation
\begin{equation} 
\label{eq:measure}
\rmd\mu_{\beta}(S) = \rmd\mu_{\beta}(U' S V'),
\end{equation}
which is assumed to be normalized, i.e.,
\begin{equation} 
\int \rmd\mu_{\beta}(S) = 1.
\end{equation}
The matrices $U'$ and $V'$ are arbitrary but fixed $n\times n$ unitary matrices for $\beta=2$, $V'={U'}^T$ for $\beta=1$, and  $V'={U'}^R$ for $\beta=4$.  \Eref{eq:measure} defines the Circular Ensembles, Orthogonal ($\beta=1$), Unitary ($\beta=2$), and Symplectic ($\beta=4$).  
For these ensembles the average of the scattering matrix satisfies $\langle S\rangle=0$.

When direct processes due to short trajectories exist (prompt responses), the $S$ matrix is no longer uniformly distributed. It turns out that in this case the direct processes can be characterized by the average  $\langle S\rangle (\neq 0)$, known as the optical $S$-matrix. In fact, in the  maximum-entropy approach developed in the past \cite{pier-book,MPS,friedman-mello}, the probability distribution of $S$ is given by
\numparts
\begin{equation}
\label{eq:dP}
\rmd P^{(\beta )}_{\langle S\rangle}(S)
=p^{(\beta )}_{\langle S\rangle}(S)\, \rmd \mu _{\beta }(S) ,
\end{equation}
where 
\begin{equation}
\label{eq:poisson}
p^{(\beta)}_{\left\langle S\right\rangle }(S)
= \frac
{[{\rm det}(I_n-\left\langle S\right\rangle \left\langle S\right\rangle ^{\dag})]^{(\beta n+2-\beta)/2}}
{|{\rm det}(I_n-S\left\langle S\right\rangle ^{\dag})|^{(\beta n+2-\beta)}}
\end{equation}
\endnumparts
is the so-called Poisson kernel.  
We note that for $\langle S \rangle=0$ the Poisson kernel reduces to unity. Therefore, it is clear that the presence of prompt scattering processes makes the analysis of the 
statistical properties of $S$  more difficult. Fortunately, a scattering matrix which satisfies the Poisson distribution, Equations~(\ref{eq:dP}) and~(\ref{eq:poisson}), can be transformed into a matrix $S_0$ with uniform distribution by an {\it ad hoc} transformation which maps the problem with presence of direct processes into one without such processes, namely 
\begin{equation}
\label{eq:transformation}
S_0 = \frac 1{t'_c} ( S - \langle S \rangle )
\frac 1{I_n - \langle S\rangle^{\dagger} S} t_c^{\dagger},
\end{equation}
where $t_c$ and $t'_c$ satisfy \cite{wrmello, brouwer-model}
\numparts
\begin{eqnarray}
\label{eq:tp}
t_c^{\dagger} t_c & = & I_n - 
\langle S\rangle^{\dagger}\langle S\rangle, \\
\label{eq:tpp}
t'_c{t'_c}^{\dagger} & = & I_n - 
\langle S\rangle\langle S\rangle^{\dagger}.
\end{eqnarray}
\endnumparts
It has been shown that for one-energy scattering matrices, the Jacobian $J_{\beta}$ of the transformation given in \eref{eq:transformation} is just the Poisson kernel, 
\eref{eq:poisson}~\cite{pier-book,friedman-mello,hua,wrmello}, i.e., 
\begin{equation}
\label{eq:Jacobian}
\rmd\mu_{\beta}(S_0)= J_{\beta} \, \rmd\mu_{\beta}(S) \equiv 
\frac
{[{\rm det}(I_n - \left\langle S\right\rangle
\left\langle S\right\rangle ^{\dag})]^{(\beta n+2-\beta )/2}}
{|{\rm det}(I_n-S\left\langle S\right\rangle^{\dag})|^{\beta n+2-\beta}}\,
\rmd\mu_{\beta}(S).
\end{equation}
The transformation \eref{eq:transformation} was also used in Ref.~\cite{gopar-mello-reduction} to show that the statistical properties of the transformed unitary scattering matrices 
at several energies $S_0(E_1), S_0(E_2), ...$ are the ones associated to the
problem of scattering in the absence of direct processes.

We note from \eref{eq:transformation} that for $\langle S\rangle=0$, $S$ reduces to $S_0$ and a uniform distribution is recovered for $S$, i.e.,  
$\rmd P^{(\beta )}_{\langle S\rangle}(S)$ reduces to the invariant measure $\rmd\mu _{\beta }(S)$. 
Thus, \eref{eq:Jacobian} implies that {\em if $S_0$ is uniformly distributed, then the matrix $S$ obtained from \eref{eq:transformation} is distributed according to the Poisson kernel}.

Although the Poisson kernel distribution has been successfully verified, for example in describing several transport properties in quantum dots~\cite{pier-book,beenakker-review,wrmello,evgeny,kottos,fyodorov2,fyodorov3} and disordered waveguides~\cite{MMAMS}, there are situations where flux conservation is violated. Therefore, the corresponding $S$ matrix becomes non-unitary, i.e., \eref{eq:flux} is not satisfied. For instance, power losses (absorption) are  unavoidable in experiments on microwave cavities and graphs~\cite{Alt,Barth,Sirko}, acoustic 
resonators~\cite{Schaadt}, and elastic media~\cite{Morales}; in those cases the $S$ matrix is sub-unitary. Also, systems with gains (amplification) exist~\cite{b-p-b}, where $S$ is a supra-unitary matrix. 

Recently, several efforts have been made in order to incorporate the 
information of losses or gains (for a review see~\cite{Fyodorov} and \cite{kuhl2}; see also~\cite{ b-p-b,MMMS,Anlage,moises-pier,dominguez,baez}). However, few investigations have considered non-unitary scattering matrices ${\tilde S}$ in the presence of direct processes. For example, sub-unitary scattering matrices have been considered in 
Refs.~\cite{kuhl,mendez,brouwer-beenakker}. There, the prompt responses come from direct reflections due to an imperfect coupling of the antenna to the cavity and the optical scattering matrix $\langle{\tilde S}\rangle$ was assumed to be a diagonal matrix. In what follows we will consider the general case when $\langle{\tilde S}\rangle$ is a full matrix.

With the same philosophy of Ref.~\cite{friedman-mello}, in this paper we transform a non-unitary scattering matrix ${\tilde S}$, with 
$\langle {\tilde S}\rangle\neq 0$, into a non-unitary scattering matrix 
${\tilde S}_0$ for which $\langle{\tilde S}_0\rangle=0$, by using the transformation  \eref{eq:transformation}. We show that the Jacobian ${\tilde J_\beta}$ associated to this  transformation  is given by the square of the Poisson kernel, \eref{eq:poisson} with $S$ replaced by ${\tilde S}$. We consider the cases of $n\times n$ non-unitary scattering matrices in the presence ($\beta=1$) and absence ($\beta=2$) of time reversal symmetry. Self dual ${\tilde S}$-matrices ($\beta=4$) are also studied.

The paper is organized as follows. The next section is devoted to the invariant measure for non-unitary matrices. In Sect.~\ref{sec:reduction} we write ${\tilde S}_0$ in terms of ${\tilde S}$ using the transformation in  \eref{eq:transformation} and the corresponding Jacobian ${\tilde J_\beta}$ is calculated. The one-channel case is presented as a simple example in Sect.~\ref{sec:n=1}. Finally, we present the conclusions in Sect.~\ref{sec:conclusions}.

\section{The invariant measure for non-unitary matrices}

Let ${\tilde S}$ be an $n\times n$  non-unitary scattering matrix, 
\begin{equation}
{\tilde S}{\tilde S}^{\dagger} \neq I_n .
\end{equation}
As for the unitary scattering matrices, we consider the following cases. In the presence of time-reversal invariance $\tilde S=\tilde S^T$ and 
we will refer to this case as $\beta=1$; in the absence of time-reversal invariance  $\tilde S$ has no restriction, and we denote this case by $\beta=2$. Finally, when $\tilde S=\tilde S^R$, $\beta=4$.

As in \eref{eq:paraS}, any non-unitary matrix $\tilde S$ can be  parametrized as \cite{hua,brouwer-beenakker,Kogan}
\begin{equation}
\label{eq:paratS}
{\tilde S} = U\brho V, 
\end{equation}
where $\brho$ is a real diagonal matrix whose diagonal elements are positive, $\brho_{ab}=\rho_a\delta_{ab}$. For sub-unitary matrices the eigenvalues of ${\tilde S}{\tilde S}^{\dagger}$ are real numbers between 0 and 1, while they are larger than 1 for supra-unitary matrices. 
$U$ and $V$ are defined in the same way than in \eref{eq:paraS} for 
each $\beta$ symmetry. Note that for  $\rho_a=1$, ${\tilde S}$ reduces to a unitary $S$-matrix and ${\tilde S}$ vanishes for $\rho_a=0$. In this sense, the diagonal elements $\rho_a$ determine the strength of the absorption or amplification~\cite{brouwer-beenakker}. 

The uniform distribution for an ensemble of ${\tilde S}$-matrices is given by the invariant measure, defined as in \eref{eq:measure}, by replacing $S$ by ${\tilde S}$. In terms of the independent elements of $\tilde S$, $\rmd\mu_{\beta}({\tilde S})$ is given 
by~\cite{metha-book,friedman-mello}
\begin{equation}
\label{eq:dmutS-1}
\rmd\mu_{\beta}({\tilde S}) = 
\prod_{\{a,b\}} \rmd (\mathrm{Re} 
{\tilde S}_{ab}) 
\rmd (\mathrm{Im} {\tilde S}_{ab}),  \ \  \mbox{for $\beta=1, 2$}.
\end{equation}
We have used the following notation for the indices for symmetric complex ($\beta=1$) and complex $(\beta=2)$ $\tilde S$-matrices:
\begin{equation}
\label{eq:indexab}
\{a,b\} =\left\{
\begin{array}{ll}
a\leq b=1,\ldots, n & \mbox{for $\beta=1$},\\
a, b =1,\ldots, n & \mbox{for $\beta=2$}.\\
\end{array}
\right.
\end{equation}
For self-dual $\tilde S$-matrices ($\beta=4$) with complex quaternion components $\tilde S_{ab}^{(\alpha)}$ we have  
\begin{equation}
\label{eq:dmutS-4}
\rmd\mu_{4}({\tilde S}) =\prod_{a < b} \prod_{\alpha=1}^3 \rmd \left( \mathrm{Re}\tilde S_{ab}^{(\alpha)}\right) \rmd \left(\mathrm{Im}\tilde S_{ab}^{(\alpha)}\right) \prod_{a \le b} \rmd \left( \mathrm{Re}\tilde S_{ab}^{(0)}\right) \rmd \left(\mathrm{Im}\tilde S_{ab}^{(0)}\right) .
\end{equation}
Equations~(\ref{eq:dmutS-1}) and~(\ref{eq:dmutS-4}) ensure that $\langle{\tilde S}\rangle=0$, as it should be for a uniform distribution of ${\tilde S}$. 

By differentiation of \eref{eq:paratS} we obtain
\begin{equation}
\label{eq:dtS}
\rmd {\tilde S} = \left( U \brho^{1/2}\right)\, \delta M\, 
\left( \brho^{1/2} V\right),
\end{equation}
where 
\begin{equation}
\label{eq:deltaM}
\delta M = 
\brho^{-1/2}\, U^{\dagger}\, dU\, \brho^{1/2} +
\brho^{-1/2}\,d\brho\,\,\brho^{-1/2} +
\brho^{1/2}\, dV\,\, V^{\dagger}\,\brho^{-1/2}.
\end{equation}
We note that $\delta M$ is in general a complex matrix ($\beta=2$); in addition, it is symmetric for $\beta=1$ and self-dual for $\beta=4$. Alternatively, the invariant measure can be written as  
\begin{equation}
\label{eq:dmutS-2}
\rmd\mu_{\beta}({\tilde S}) = 
\prod_{\{a,b\}}  \mathrm{Re}( 
{\delta M}_{ab}) \mathrm{Im}( {\delta M}_{ab}), \ \  \ \ \mbox{for $\beta=1, 2$}, 
\end{equation}
and, for $\beta=4$,
\begin{equation}
\label{eq:dmutS-4b}
\rmd\mu_{4}({\tilde S}) =
\prod_{a < b} \prod_{\alpha=1}^3  
\mathrm{Re}({\delta M}_{ab}^{(\alpha)})
\mathrm{Im}({\delta M}_{ab}^{(\alpha)})  
\prod_{a \le b}  
\mathrm{Re} ({\delta M}_{ab}^{(0)}) 
\mathrm{Im}({\delta M}_{ab}^{(0)}) ,
\end{equation}
where $\delta M_{ab}^{(\alpha)}$ is the $\alpha$-th quaternion component of 
$\delta M_{ab}$ (see \ref{app}).


\section{Mapping the scattering problem from presence to absence of direct processes}
\label{sec:reduction}

Let us consider an ensemble of non-uniformly distributed ${\tilde S}$ matrices with average  $\langle{\tilde S}\rangle\ne 0$. We now apply the transformation given by \eref{eq:transformation} to $\tilde S $ matrices in order to obtain 
an ensemble of $\tilde S_0$ matrices with uniform distribution, i.e.,
\begin{equation}
\label{eq:trans-2}
{\tilde S}_0 = \frac 1{\tilde t'_c} 
\left( {\tilde S} - \langle{\tilde S} \rangle \right)
\frac 1{I_n - \langle{\tilde S}\rangle^{\dagger}{\tilde S}} {\tilde t_c}^{\dagger},
\end{equation}
where $\tilde t_c$ and $\tilde t'_c$ satisfy (similarly to  Equations~(\ref{eq:tp}) and~(\ref{eq:tpp}))
\numparts
\begin{eqnarray}
\label{eq:tildetp}
\tilde t_c^{\dagger} \tilde t_c & = & I_n - 
\langle {\tilde S}\rangle^{\dagger}\langle {\tilde S}\rangle, \\
\label{eq:tildetpp}
\tilde t'_c {\tilde{t}_c}'^{\dagger} & = & I_n - 
\langle {\tilde S}\rangle\langle {\tilde S}\rangle^{\dagger}.
\end{eqnarray}
\endnumparts
Then we look for the Jacobian ${\tilde J_\beta}$ of this transformation.

By direct differentiation of ${\tilde S}_0$ we obtain 
\begin{equation}
\label{eq:dS0}
\rmd {\tilde S}_0 = A'\, \rmd{\tilde S}\, A,
\end{equation}
with $A$ and $A'$ the complex matrices
\numparts
\begin{eqnarray}
\label{eq:A}
A & = & \left( I_n - 
\langle{\tilde S}\rangle^{\dagger}
{\tilde S}\right)^{-1} {\tilde t}_c^{\dagger} \, ,\\
\label{eq:Ap}
A' & = & \tilde{t'}_c^{\dagger} \left( I_n - {\tilde S}
\langle{\tilde S}\rangle^{\dagger}
\right)^{-1},
\end{eqnarray}
\endnumparts
for $\beta=2$. $A'=A^T$ for $\beta=1$ and $A'=A^R$ for $\beta=4$. On the other hand, a similar expression to \eref{eq:dtS} is valid for ${\tilde S}_0$ as well, namely  
\begin{equation}
\label{eq:dtS0}
\rmd {\tilde S_0} = \left( U_0 \brho_0^{1/2}\right)\, \delta M_0\, 
\left( \brho_0^{1/2} V_0\right).
\end{equation}
Substituting Equations~(\ref{eq:dtS}) and~(\ref{eq:dtS0}) into \eref{eq:dS0}, we obtain the relation between the matrices $\delta M_0$ and  $\delta M$, 
\begin{equation}
\label{deltaM0}
\delta M_0 = B'\, \delta M \, B ,
\end{equation}
where $\delta M_0$ has the same structure of \eref{eq:deltaM} and 
\numparts
\begin{eqnarray}
\label{eq:B}
B & = & \brho^{1/2} VAV_0^{-1}\brho_0^{-1/2} ,\\
\label{eq:Bp}
B' & = & \brho_0^{-1/2} U_0^{-1}A'U\brho^{1/2} ,
\end{eqnarray}
\endnumparts
for $\beta=2$. Again, $B'=B^T$ for $\beta=1$ and $B'=B^R$ for $\beta=4$. We remark that $\delta M_0$ and $\delta M$ are complex matrices, hence $B$ and $B'$ are also complex, in contrast to the case of unitary scattering matrices where $\delta M_0$ and $\delta M$ may be taken as real matrices. 
In the following we will restrict ourselves to the cases $\beta=1$ and 2, unless explicitly indicated otherwise. The case $\beta=4$ is considered in \ref{app}.

It is convenient to separate the real and imaginary parts of $\delta M_0$ and $\delta M$ to obtain the Jacobian of the transformation as in \eref{eq:dmutS-2}, 
\numparts
\begin{eqnarray}
\label{eq:real}
\!\!\!\!\!\!\!\!\!\!\!\!\mathrm{Re}(\delta{M_0}_{ab}) & = &
\sum_{c,d=1}^n\mathrm{Re}(B'_{ac}B_{db}) \mathrm{Re}(\delta M_{cd}) +
\sum_{c,d=1}^n\mathrm{Re}(\rmi B'_{ac}B_{db}) \mathrm{Im}(\delta M_{cd}),
\\
\label{eq:imaginary}
\!\!\!\!\!\!\!\!\!\!\!\!\mathrm{Im}(\delta{M_0}_{ab}) & = &
\sum_{c,d=1}^n\mathrm{Im}(B'_{ac}B_{db}) \mathrm{Re}(\delta M_{cd}) +
\sum_{c,d=1}^n\mathrm{Im}(\rmi B'_{ac}B_{db}) \mathrm{Im}(\delta M_{cd}). 
\end{eqnarray}
\endnumparts
Next, we calculate the Jacobian ${\tilde J}^{(\beta)}_{ab}$ of the transformation which relates the real and imaginary parts of the independent elements of $\delta M_0$ with those of $\delta M$ as 
\begin{equation}
\label{eq:Jabdefinition}
\mathrm{Re}(\delta{M_0}_{ab})\mathrm{Im}(\delta{M_0}_{ab}) = 
{\tilde J}^{(\beta)}_{ab} \,
\mathrm{Re}(\delta{M}_{ab})\mathrm{Im}(\delta{M}_{ab}).
\end{equation}
To this end, and following Ref.~\cite{friedman-mello}, we start by assuming that $B$ and $B'$ are diagonal matrices, before considering the general case of any complex matrix.


\subsection{\label{diagonal} A simple example: $B$ and $B'$ diagonal matrices}

Let $B$ and $B'$ 
\numparts
\begin{eqnarray}
B_{ab} & = & \lambda_a\, \delta_{ab} , \\
B'_{ab} & = & \lambda'_{a}\, \delta_{ab},
\end{eqnarray}
\endnumparts
where $\lambda_a$'s and $\lambda'_a$'s are complex numbers. 
Therefore \eref{eq:real} and~\eref{eq:imaginary} reduce to
\numparts
\begin{eqnarray}
\label{eq:real-2}
\mathrm{Re}(\delta{M_0}_{ab}) & = &
\mathrm{Re}(\lambda'_a\lambda_b)\, \mathrm{Re}(\delta M_{ab}) -
\mathrm{Im}(\lambda'_a\lambda_b)\, \mathrm{Im}(\delta M_{ab}), \\
\label{eq:imaginary-2}
\mathrm{Im}(\delta{M_0}_{ab}) & = &
\mathrm{Im}(\lambda'_a\lambda_b)\, \mathrm{Re}(\delta M_{ab}) +
\mathrm{Re}(\lambda'_a\lambda_b)\, \mathrm{Im}(\delta M_{ab}).
\end{eqnarray}
\endnumparts
From these two equations, the Jacobian ${\tilde J}^{(\beta)}_{ab}$ is given by
\begin{equation}
\label{eq:Jab}
{\tilde J}^{(\beta)}_{ab} =  [\mathrm{Re}(\lambda'_a\lambda_b)]^2 + [\mathrm{Im}(\lambda'_a\lambda_b)]^2 = 
|\lambda'_a\lambda_b|^2.
\end{equation}

\subsection{\label{general}The general case}

Any complex matrices $B$ and $B'$ 
can be written as 
\numparts
\begin{eqnarray}
\label{eq:B-2}
B & = & Q \Lambda O ,\\
\label{eq:Bp-2}
B' & = & Q'\Lambda' O',
\end{eqnarray}
\endnumparts
where $Q$, $Q'$, $O$, $O'$ are the most general unitary matrices; we recall that for $\beta=1$, $B'=B^T$, hence $Q'=O^T$ and $O'=Q^T$. Here, $\Lambda$ and $\Lambda'$ are diagonal matrices whose diagonal elements 
$\lambda_a$ and $\lambda_a'$ are complex numbers. 
We substitute \eref{eq:B-2} and~\eref{eq:Bp-2} into 
\eref{deltaM0} to obtain
\begin{equation}
\label{eq:dM0}
\delta M_0 = Q' \delta M_2 O,
\end{equation}
where we have defined 
\begin{eqnarray}
\label{eq:dM2}
\delta M_2=\Lambda'\delta M_1\Lambda, \\
\label{eq:dM1}
\delta M_1=O'\delta M Q.
\end{eqnarray} 

We now calculate the Jacobian of the transformation 
$\delta M_1\to\delta M_2$ given by \eref{eq:dM2}. As before, it is convenient to separate the real and imaginary parts, 
\numparts
\begin{eqnarray}
\label{eq:dM2real}
\mathrm{Re}(\delta{M_2}_{ab}) & = &
\mathrm{Re}(\lambda'_a\lambda_b)\, \mathrm{Re}(\delta{M_1}_{ab}) - 
\mathrm{Im}(\lambda'_a\lambda_b)\, \mathrm{Im}(\delta{M_1}_{ab}) , 
\\
\label{eq:dM2imaginary}
\mathrm{Im}(\delta{M_2}_{ab}) & = &
\mathrm{Im}(\lambda'_a\lambda_b)\, \mathrm{Re}(\delta{M_1}_{ab}) +
\mathrm{Re}(\lambda'_a\lambda_b)\, \mathrm{Im}(\delta{M_1}_{ab}). 
\end{eqnarray}
\endnumparts
This is exactly the same transformation as in \eref{eq:real-2} and~\eref{eq:imaginary-2} for the case of diagonal matrices $B$ and $B'$. Then, the corresponding Jacobian 
of the transformation in \eref{eq:dM2real} and~\eref{eq:dM2imaginary} 
is ${\tilde J}_{ab}^{(\beta)}$ given by \eref{eq:Jab}, from which we obtain
\begin{equation}
\label{eq:dM2-dM1}
\prod_{\{a,b\}} 
\mathrm{Re}(\delta{M_2}_{ab})
\mathrm{Im}(\delta{M_2}_{ab})
= {\tilde J}_{\beta}
\prod_{\{a,b\}} 
\mathrm{Re}(\delta{M_1}_{ab})
\mathrm{Im}(\delta{M_1}_{ab}),
\end{equation}
where 
\begin{equation}
\label{eq:Jresult-1}
{\tilde J}_{\beta} = \prod_{\{a,b\}} {\tilde J}_{ab}^{(\beta)} = 
\left|\prod_{\{a,b\}}\lambda'_a\lambda_b\right|^2.
\end{equation}
In~\ref{explicit} we show the explicit calculation of ${\tilde J}_{\beta}$. 
We note, on the other hand, that the transformation $\delta M\to Q'\delta MO$ does not change the 
measure~\cite{metha-book}; therefore, from  \eref{eq:dM0} and~\eref{eq:dM1}, $\delta M_0$ has the same measure as $\delta M_2$ while $\delta M_1$ has the same measure as $\delta M$. Thus, by transitivity, 
\begin{equation}
\label{eq:dM0-dM}
\prod_{\{a,b\}} 
\mathrm{Re}(\delta{M_0}_{ab})
\mathrm{Im}(\delta{M_0}_{ab})
= {\tilde J}_{\beta}
\prod_{\{a,b\}} 
\mathrm{Re}(\delta{M}_{ab})
\mathrm{Im}(\delta{M}_{ab}),
\end{equation}
which means that the Jacobian ${\tilde J}_{\beta}$ of the transformation given in \eref{eq:trans-2} relates the invariant measures $d\mu({\tilde S}_0)$ and $d\mu({\tilde S})$. Inserting the last result in \eref{eq:dmutS-2} we obtain
\begin{equation}
\label{eq:dS0-dS}
d\mu({\tilde S}_0) = {\tilde J}_{\beta} \, d\mu({\tilde S}).
\end{equation}

From 
\eref{eq:Jresult-1} for $\beta=1$ and 2, and~(\ref{J4}) for $\beta=4$, ${\tilde J}_{\beta}$ can be written in a single expression as 
\begin{equation}
\label{eq:Jresult-2}
{\tilde J}_{\beta} = \left|(\det B')^{(\beta n+2-\beta)/2}
(\det B)^{(\beta n+2-\beta)/2}\right|^2.
\end{equation}
Also \eref{eq:B}  and \eref{eq:Bp} imply that $\det B'=\det A'$ and $\det B=\det A$; hence, ${\tilde J}_{\beta}$ can be expressed as 
\begin{equation}
\label{eq:Jresult-3}
{\tilde J}_{\beta} =
\left|\det \left( A'A\right)\right|^{(\beta n+2-\beta)}.
\end{equation}
Using \eref{eq:A} and \eref{eq:Ap} in the last expression we have 
\begin{eqnarray}
\det (A'A) = 
\frac{\det {\tilde{t'}_c}^{\dagger}\det \tilde{t}_c^{\dagger} }
{\det \left( I_n - {\tilde S}\left\langle{\tilde S}\right\rangle\right)^2}.
\end{eqnarray}

Finally, from \eref{eq:tildetp} and \eref{eq:tildetpp} we can verify that $|\det \tilde t'_c|=|\det \tilde t_c|$. Then 
${\tilde J}_{\beta}$ is given by 
\begin{equation}
\label{eq:Jresult-4}
{\tilde J}_{\beta} = \left[
\frac{\left| \det \left( I_n - 
\left\langle{\tilde S}\right\rangle
\left\langle{\tilde S}\right\rangle^{\dagger}\right)
\right|^{(\beta n+2-\beta)/2} }
{\left| \det \left( I_n - 
{\tilde S}\left\langle{\tilde S}\right\rangle^{\dagger}\right)
\right|^{(\beta n+2-\beta)}}\right]^2.
\end{equation}

\Eref{eq:Jresult-4}, together with \eref{eq:dS0-dS}, is the main result of this work and can be interpreted as follows: {\em if a non-unitary scattering matrix ${\tilde S}_0$ is uniformly distributed in the space of non-unitary scattering matrices, then another non-unitary scattering matrix ${\tilde S}$, obtained from ${\tilde S}_0$ through the transformation given by \eref{eq:trans-2}, is distributed according to 
${\tilde J}_{\beta}$ given by \eref{eq:Jresult-4}}. In this sense, ${\tilde J}_{\beta}$ is the generalization of the Poisson kernel for non-unitary scattering matrices. We show in \ref{app2} that \eref{eq:dS0-dS} together with \eref{eq:Jresult-4} yields the Poisson kernel for unitary matrices. 
We remark that for unitary scattering matrices, the Poisson kernel has been originally obtained in the framework of maximum-entropy or Shannon information theory \cite{pier-book}. However, to the best of our knowledge, a derivation of the Poisson kernel for non-unitary scattering  matrices from maximum-entropy arguments is not available.


\section{The one-channel case}
\label{sec:n=1}

As an example, let us consider a $1\times 1$ matrix 
${\tilde S}$ which can be parametrized in polar form as 
\begin{equation}
\label{eq:Sn=1}
{\tilde S} = \sqrt{R}\, \rme^{\rmi\theta},
\end{equation}
where $R$ is the reflection coefficient and $\theta$ is the negative of twice the phase shift with $0\leq R<1$ for sub-unitary matrices and $R>1$ for supra-unitary matrices. 

In the sub-unitary case, a uniform distribution for 
${\tilde S}$ means that it is distributed according to its invariant measure
\begin{equation}
\label{eq:dmun=1}
\rmd\mu_{\beta}({\tilde S}) = \rmd R \,\frac{\rmd\theta}{2\pi}. 
\end{equation}
A non-uniform distribution of ${\tilde S}$ is constructed from \eref{eq:dmun=1} as 
\begin{equation}
\label{eq:dPn=1}
\rmd P^{(\beta)}({\tilde S}) = 
p^{(\beta)}(R,\theta) \rmd R \,\frac{\rmd\theta}{2\pi} .
\end{equation}
For instance, if ${\tilde S}_0$ is the scattering matrix associated to chaotic cavities with losses in the absence of direct processes, $p_0^{(\beta)}(R_0,\theta_0)=p_0^{(\beta)}(R_0)$, where $p_0^{(\beta)}(R_0)$
is known \cite{Fyodorov,MMMS}. In the presence of direct processes, according to \eref{eq:dS0-dS}, with ${\tilde J}_{\beta}$ given by \eref{eq:Jresult-4}, we have
\begin{equation}
\rmd R_0\,\frac{\rmd\theta_0}{2\pi} = \left(
\frac{1 - \left|\left\langle{\tilde S}\right\rangle\right|^2}
{\left| 1 - {\tilde S}\left\langle{\tilde S}\right\rangle^*\right|^2}
\right)^2
\rmd R\,\frac{\rmd\theta}{2\pi} ,
\end{equation}
which, multiplying by $p_0(R_0(R,\theta))$, can be written as
\begin{equation}
\label{p0R0}
p_0(R_0(R,\theta))\rmd R_0\,\frac{\rmd\theta_0}{2\pi} = \left(
\frac{1 - \left|\left\langle{\tilde S}\right\rangle\right|^2}
{\left| 1 - {\tilde S}\left\langle{\tilde S}\right\rangle^*\right|^2}
\right)^2 p_0(R_0(R,\theta))
\rmd R\,\frac{\rmd\theta}{2\pi} .
\end{equation}
Comparing the right hand of side of Equations~(\ref{eq:dPn=1}) and~(\ref{p0R0}), we obtain that 
\begin{equation}
\label{eq:pn=1}
p^{(\beta)}(R,\theta) = 
\left(
\frac{1 - \left|\left\langle{\tilde S}\right\rangle\right|^2}
{\left| 1 - {\tilde S}\left\langle{\tilde S}\right\rangle^*\right|^2}
\right)^2 p_0^{(\beta)}(R_0(R,\theta)) .
\end{equation}
This result, is in agreement with that of Ref.~\cite{kuhl}, where $p^{(1)}(R,\theta)$ was verified by comparing with experimental measurements in microwave chaotic cavities.

\section{Conclusions}
\label{sec:conclusions}

We have reduced the problem of scattering in the presence of direct processes to the case without such processes for $n \times n$ {\it non-unitary} scattering matrices. We use a transformation to map an ensemble of  
such matrices ${\tilde S}$ with $\langle \tilde S \rangle \ne 0$ 
to an ensemble of $\tilde S_0$ scattering matrices with $\langle \tilde S_0 \rangle = 0$. In our theoretical framework, the direct processes are characterized by the average $\langle \tilde S \rangle$. Therefore, $\tilde S $ and $\tilde S_0 $ describe a system in the presence
and in the absence of direct processes, respectively. 
The Jacobian  ${\tilde J}_{\beta}$ of the transformation turns out to be the square of  the known Poisson kernel. In our analysis 
we consider general complex, symmetric, and self-dual scattering matrices, in analogy to the three basic symmetries in Dyson's scheme $\beta=2,1,$ and 4. 
We have found that if ${\tilde S}_0$ is uniformly distributed in the space of non-unitary scattering matrices, then ${\tilde S}$ obtained from ${\tilde S}_0$ through the transformation (\ref{eq:trans-2}) is distributed according to 
${\tilde J_\beta}$ given in \eref{eq:Jresult-4}. Our study extends known results for the probability distribution of non-unitary scattering matrices in the absence of direct process to the case in the presence of such processes. As a consequence, from the simplest case of a uniform distribution of $\tilde S$, it is possible to obtain more complex distributions, emerging from situations where prompt responses are relevant to the scattering problem, as has been illustrated in the one-channel case.


\appendix

\section{Jacobian ${\tilde J}_{\beta}$ for $\beta=4$}
\label{app}

The invariant measure for $\beta=4$ is given by \eref{eq:dmutS-4b}. For  self-dual $\delta M_0$ and $\delta M$ matrices, \eref{deltaM0} implies that 
\begin{equation}
B'=B^R .
\end{equation}
Using the standard notation for quaternions~\cite{metha-book}, the 
elements of $\delta M$ can be written as 
\begin{equation}
\delta M_{ab}= \sum_{\alpha=0}^3 \delta M_{ab}^{(\alpha)}e_{\alpha}, \qquad \alpha=0,1,2,3,
\end{equation}
where $\delta M_{ab}^{(\alpha)}$ is the projection of $\delta M_{ab}$ on the quaternion $e_{\alpha}$, where 
\begin{equation}
e_0=\left[\begin{array}{cc} 1 & 0 \\ 0 & 1  \end{array} \right], \, 
e_1=\left[\begin{array}{cc} i & 0 \\ 0 & -i \end{array} \right], \, 
e_2=\left[\begin{array}{cc} 0 & 1 \\ -1 & 0 \end{array} \right], \,
e_3=\left[\begin{array}{cc} 0 & i \\ i & 0  \end{array} \right] .
\end{equation}
From \eref{deltaM0}, we have
\begin{equation}
\label{deltaM0ij}
\delta {M_0}_{ij}^{(\alpha)}=
\sum_{a,b}B'_{ia} \delta M_{ab}^{(\alpha)} B_{bj}.
\end{equation}
As in Sect.~\ref{diagonal}, we consider first the special case of a diagonal matrix $B$ with
\begin{eqnarray}
\label{bdiagonal}
B_{a b} &=& \lambda_a \delta_{ab} ,\\
\label{bpdiagonal}
B'_{a b} &=& \lambda'_a \delta_{ab}.
\end{eqnarray}
Substituting Equations~(\ref{bdiagonal}) and~(\ref{bpdiagonal}) into \eref{deltaM0ij} we obtain
\begin{equation}
\delta {M_0}_{ij}^{(\alpha)}=\lambda'_i \lambda_j \delta M^{(\alpha)}_{ij}.
\end{equation}
Hence, the real and imaginary parts are related by
\begin{eqnarray}
\label{eq:ri1}
\mathrm{Re}(\delta {M_0}^{(\alpha)}_{ij})&=&\mathrm{Re}(\lambda'_i \lambda_j)\mathrm{Re}(\delta {M}^{(\alpha)}_{ij})- \mathrm{Im}(\lambda'_i \lambda_j)\mathrm{Im}(\delta {M}^{(\alpha)}_{ij}) \\
\label{eq:ri2}
\mathrm{Im}(\delta {M_0}^{(\alpha)}_{ij})&=&\mathrm{Im}(\lambda'_i \lambda_j)\mathrm{Re}(\delta {M}^{(\alpha)}_{ij})+ \mathrm{Re}(\lambda'_i \lambda_j)\mathrm{Im}(\delta {M}^{(\alpha)}_{ij}) .
\end{eqnarray}
Therefore, we have
\begin{equation}
\mathrm{Re}(\delta {M_0}^{(\alpha)}_{ij})\mathrm{Im}(\delta {M_0}^{(\alpha)}_{ij})=|\lambda'_i \lambda_j|^2\mathrm{Re}(\delta M_{ij}^{(\alpha)})\mathrm{Im}(M_{ij}^{(\alpha)}),
\end{equation}
where $|\lambda'_i \lambda_j|^2$ is just the Jacobian of the transformation given by Equations~(\ref{eq:ri1}) and~(\ref{eq:ri2}). Thus, using \eref{eq:dmutS-4b} we obtain
\begin{equation}
\mathrm{d}\mu_4({\tilde S}_0)={\tilde J}_4 
\mathrm{d}\mu_4(\tilde S),
\end{equation}
where 
\begin{eqnarray}
{\tilde J}_4 & = & 
\prod_{\alpha=1}^3 \prod_{i<j}|\lambda'_i \lambda_j|^2 
\prod_{i\le j}|\lambda'_i \lambda_j|^2 = 
\left| \frac{\left( \prod_{i\le j}\lambda_i' \lambda_j \right)^4}
{(\prod_i \lambda'_i \lambda_i)^3} \right|^2 
\nonumber \\ \label{J4} & = & 
\left|\frac{(\mathrm{det}B')^{2(n+1)} 
(\mathrm{det}B)^{2(n+1)}}{(\mathrm{det}B')^{3} 
(\mathrm{det}B)^{3}} \right|^2 
= \left| (\mathrm{det}B^R)^{2n-1} (\mathrm{det}B)^{2n-1} \right|^2,
\end{eqnarray}
which is the result presented in \eref{eq:Jresult-2} for $\beta=4$. 

As in Sect.~\ref{general}, the general case of a non diagonal matrix $B$ can be reduced to a diagonal one by means of Equations~(\ref{eq:B-2}) and~(\ref{eq:Bp-2}) with $Q'=O^R$ and $O'=Q^R$. Thus, we have the same result for ${\tilde J}_4$. 


\section{Explicit calculation of Eq. (\ref{eq:Jresult-1})}
\label{explicit}

For $\beta=1$, $\lambda_a'=\lambda_a$ and the products appearing in the right-hand side of Eq. (\ref{eq:Jresult-1}) can be written explicitly as
\begin{eqnarray}
\prod_{a\leq b} \lambda_a\lambda_b & & = 
\prod_{a=1}^n \prod_{b=a}^n \lambda_a\lambda_b 
\nonumber \\ & & 
\begin{array}{rcr}
=(\lambda_1\lambda_1) (\lambda_1\lambda_2)(\lambda_1\lambda_3) &
\cdots & (\lambda_1\lambda_{n-1})(\lambda_1\lambda_n) 
\\ \times \qquad\quad 
(\lambda_2\lambda_2)(\lambda_2\lambda_3) & \cdots &  (\lambda_2\lambda_{n-1})(\lambda_2\lambda_n)
\\ \times \qquad\qquad\quad\,\,\,\, (\lambda_3\lambda_3) & \cdots &  (\lambda_3\lambda_{n-1})(\lambda_3\lambda_n) 
\\ & & \vdots \qquad\quad
\\ \times\qquad\qquad\qquad\qquad\,\, & &
(\lambda_{n-1}\lambda_{n-1})(\lambda_{n-1}\lambda_n) 
\\ \times\qquad\qquad\qquad\qquad\,\, & & 
(\lambda_n\lambda_n) 
\end{array}
\end{eqnarray}
We can see that $\lambda_1$ appears $n+1$ times in the first line only, $\lambda_2$ once in the first line and $n$ times in the second line, $\lambda_3$ appears once in the first and second lines and $n-1$ times in the third line, etc. Then each $\lambda_a$ appears $n+1$ times in total and the product can be written as 
\begin{eqnarray}
\prod_{a\leq b} \lambda_a\lambda_b = 
\prod_{a=1}^n \lambda_a^{n+1} =
\left( \prod_{a=1}^n \lambda_a \right)^{n+1} =
(\det B)^{n+1}
\end{eqnarray}

For $\beta=2$, we have
\begin{eqnarray}
\prod_{a,b=1}^n \lambda'_a\lambda_b & = &
\prod_{a=1}^n \left( \prod_{b=1}^n \lambda'_a\lambda_b\right) =
\prod_{a=1}^n\left({\lambda'_a}^n\prod_{b=1}^n\lambda_b\right) 
\nonumber \\ & = & 
\prod_{a=1}^n{\lambda'_a}^n\prod_{b=1}^n\lambda_b^n = 
\left(\prod_{a=1}^n{\lambda'_a}\right)^n
\left(\prod_{b=1}^n\lambda_b\right)^n 
\nonumber \\ & = & 
\det {B'}^n \det B^n.
\end{eqnarray}

We can summarize the result for any $\beta$ as 
\begin{equation}
\prod_{\{a,b\}}^n \lambda'_a\lambda_b = 
\det {B'}^{(\beta n+2-\beta)/2}
\det {B}^{(\beta n+2-\beta)/2},
\end{equation}
which is also valid for $\beta=4$ (See~\ref{app}).


\section{Reduction to the Poisson kernel for unitary matrices}
\label{app2}

We show in this appendix that the original Poisson kernel can be obtained by restricting a non-unitary $\tilde S$ matrix to be unitary. 

On one hand, we are concerned with unitary scattering matrices $S_0=U_0V_0$ and $S=UV$ related by \eref{eq:transformation}. Their measures satisfy \eref{eq:Jacobian}. Therefore, if the probability density distributions of $S_0$ and $S$ are $p_0(S_0)$ and $p(S)$, respectively, $p_0(S_0)$ and $p(S)$ are related by 
\begin{equation}
\label{eq:psps0}
p(S) = p_0(S_0)\, J_{\beta}, \qquad \mbox{or} \qquad 
p_0(S_0) = \frac{p(S)}{J_{\beta}}.
\end{equation}
In particular, we assume that $p(S)=\delta(\tilde S\tilde S^{\dagger}-I_n)$, where $\tilde S=U\rho V$ is a non-unitary matrix which is a function of $S$. Then, $\tilde S_0=U_0\rho_0V_0$ is a function of $S_0$, related to $\tilde S$ via \eref{eq:trans-2}. This leads to $p_0(S_0)=\delta(\tilde S_0\tilde S_0^{\dagger}-I_n)$, where $\tilde S_0$ is a function of $S_0$. From \eref{eq:psps0} we have that
\begin{equation}
\label{eq:dd}
\delta(\rho_0-I_n) = 
\frac{\delta(\rho-I_n)}{J_{\beta}}.
\end{equation}

On the other hand, consider non-unitary scattering matrices, starting with \eref{eq:dS0-dS} with ${\tilde J}_{\beta}$ given by \eref{eq:Jresult-4}. Here, $\tilde S_0$ is uniformly distributed in the space of non-unitary scattering matrices. We restrict to the space of unitary scattering matrices, imposing unitarity on ${\tilde S}_0$ by multiplying \eref{eq:dS0-dS} by $\delta(\rho_0-I_n)$. We obtain 
\begin{equation}
\label{eq:Et-E}
\delta(\rho_0-I_n)\, d\mu({\tilde S}_0) = 
\delta(\rho_0-I_n)\, {\tilde J}_{\beta}\,
d\mu({\tilde S}).
\end{equation}
Using \eref{eq:dd}, \eref{eq:Et-E} yields
\begin{equation}
\delta(\rho_0-I_n)\, d\mu(\rho_0)\,d\mu(S_0) = 
\frac{\delta(\rho-I_n)}{J_{\beta}}\, {\tilde J}_{\beta}\,
d\mu(\rho)\, d\mu(S), 
\end{equation}
where we have written $d\mu(\tilde S)=d\mu(\rho)\,d\mu(S)$. On the left-hand side, the integral with respect to $\rho_0$ gives one, while the integral with respect to $\rho$ on the right-hand side evaluates to ${\tilde J}_{\beta}$ at $\rho=I_n$. This implies $\tilde J_{\beta}(\rho=I_n)=J_{\beta}^2$ (compare \eref{eq:Jacobian} with \eref{eq:Jresult-4}). Finally, we get
\begin{equation}
d\mu(S_0) = 
\frac{J_{\beta}^2}{J_{\beta}} \, d\mu(S) = 
J_{\beta}\, d\mu(S), 
\end{equation}
where $J_{\beta}$ is the Poisson kernel for unitary scattering matrices $S$.


\ack
We thank J. Flores and D. Sanders for carefully reading this manuscript. 
Also, we thank the DGAPA-UNAM, M\'exico for financial support, under project IN118805. VAG acknowledges financial support from the Ministerio de Educaci\'on y Ciencia, Spain, through the Ram\'on y Cajal Program and the project FIS 2006-08-532, as well as the hospitality during his visit to the ICF-UNAM and UAM-Iztapalapa. MMM thanks BIFI and the Departamento de F\'{\i}sica Te\'orica of the Universidad de Zaragoza for kind hospitality.


\end{document}